\documentclass[12pt]{article}
\usepackage{amsfonts,amssymb,subfigure}  
\usepackage{amsmath,hyperref}
\usepackage{amsthm}
\usepackage{graphicx,setspace}
\usepackage{float}
\usepackage{pdflscape}
\usepackage{geometry}
\usepackage{fancyhdr}
\usepackage[svgnames]{xcolor}
\usepackage{lineno}

\title{Modelling and prediction of the wildfire data using fractional Poisson process}
\author{Sudeep R. Bapat\footnote{Shailesh J. Mehta School of Management, Indian Institute of Technology Bombay - Powai, Mumbai 400076, Maharashtra,  INDIA. Email: sudeepb@iitb.ac.in} \and Aditya Maheshwari\footnote{ Operations Management and Quantitative Techniques Area, Indian Institute of Management Indore, Indore 453556, Madhya Pradesh, INDIA. Email: adityam@iimidr.ac.in
}}
\date{}

\begin{document}
%
\maketitle

\begin{spacing}{1}
\begin{abstract}
Modelling wildfire events has been studied in the literature using the Poisson process, which essentially assumes the independence of wildfire events. In this paper, we use the fractional Poisson process to model the wildfire occurrences in California between June 2019 - April 2023 and predict the wildfire events that explains the underlying memory between these events. We introduce  method of moments and maximum likelihood estimate approaches to estimate the parameters of the fractional Poisson process, which is an alternative to the method proposed by \cite{Cahoy2010}. We obtain the estimates of the fractional parameter as 0.8,  proving that the wildfire events are dependent. The proposed model has reduced prediction error by 90\% compared to the classical Poisson process model. 
\end{abstract}

\begin{flushleft}
\end{flushleft}

\begin{flushleft}
\end{flushleft}

\newpage

\section{Introduction}

Forest ecosystems face ongoing threats from wildfires, a pervasive ecological disturbance. Since the inception of plant communities on land, natural elements like lightning and human activities have ignited fires, posing a continual risk to forest habitats. The impact of these wildfires remains a persistent challenge in forest conservation and management. There are various mathematical model developed to study wildfires and a comprehensive summary about the efforts can be found in \cite{math-model-review}. The utilization  of Poisson count models was attempted by \cite{Poisson-1} and proved that it is superior to empirical model. Subsequent to the above paper, several researchers applied Poisson process (PP) model to study wildfire spread (see \cite{Poisson-2,Poisson-3,Poisson-4,Poisson-5,Poisson-6,Poisson-7,Poisson-8}) and establish that it is indeed an appropriate probabilistic template for wildfire prediction.  \\\\ 
However, the PP model is inadequate for capturing the bursty nature of data traffic, particularly when observed across multiple time scales. It is shown in empirical studies (see \cite{Poisson-fail2,Poisson-fail}) that the chance of network connection sessions dropping off follows a better pattern with a power law decay rather than an exponential one. Efforts to enhance or broaden the Poisson model have given rise to novel formulations like the non-homogeneous Poisson process, the Cox point process, the higher-dimensional Poisson process, and so forth. In the past decade, researchers have been particularly interested in  fractional generalizations of the Poisson process. In this paper, we propose to use the fractional Poisson process (fPP) to model wildfire events and illustrate its superiority over the classical Poisson model.  \\\\  
Repin and Saichev (2000) (see \cite{repin}) first investigated the possibility of a fractional extension to the Poisson distribution. Their investigation involved a PP where the distribution of intervals between jumps is characterized by a fractional differential equation. Jumarie (2001) (see \cite{jumarie}) provided a formal definition of the fPP through the difference-differential approach.
By using the fractional difference-differential equation technique, they were able to construct the expressions for the variance, mean, and generating functions, as well as the {\it pmf}. Furthermore, he demonstrated that the time intervals between successive occurrences of the fPP conform to the Mittag-Leffler distribution. Mainardi \textit{et. al} (2004) investigated the fPP using the renewal process methodology (see \cite{mainardi04,Mainardi2007}).  In \cite{Wang2003, Wang2006, Wang2007}, distinct methods for fractional generalization of the PP are addressed.\\\\
The fPP is better than the PP for modeling long-memory dynamics of a stream of counts in complex systems (see \cite{lask,fractionalriskprocess}). The PP assumes that the waiting times between events are exponentially distributed and independent of each other, which is not always the case in real-world systems. On the other hand, the fPP captures the dependence or long-memory effect, resulting in non-exponential waiting time distribution empirically observed in complex systems. The fPP includes an additional parameter $0<\beta\leq 1$, which makes it more flexible than the PP.\\\\ 
In \cite{Cahoy2010}, authors proposed a way to estimate the parameter of the fPP by estimating the parameters of the Mittag-Leffler inter-arrival times. In this paper, we propose two alternative ways to estimate parameters of the fPP, namely the method of moments (MoM) and the maximum likelihood estimates (MLE). Thereafter, we apply the above-mentioned new methods on wildfire data and show the superiority of the fPP model over the usual PP, while  comparing the empirical cumulative distribution functions of both processes. We also predict the occurrences of wildfires using the fPP model which reduces the prediction error by 90\% compared to the PP model.\\\\
The paper is organised as follows. Section \ref{sec:fPP} contains some preliminary definitions and results. In Section \ref{sec:mom}
and \ref{sec:MLE}, we discuss the MoM and MLE methods of estimating the parameters of the fPP. We present practical data analysis on the estimation methods in Section \ref{sec:data}. Finally, the prediction of wildfire events are presented in Section \ref{sec:pred}.

\section{Preliminaries}\label{sec:fPP}
In this section, we present some preliminaries that will be useful for the rest of the paper. 
 
\subsection{Poisson process} Let $\{N(t)\}_{t\geq0}$ be the Poisson process (PP) with rate $\lambda>0$. The probability mass function (\textit{pmf}) $p(n,t)=\mathbb{P}[N(t)=n]$ of the PP solves the following  equation
 $$ \frac{d}{dt}p(n,t) = -\lambda \left[p(n,t) - p(n-1,t)\right],$$
 with $p(n,0)=1$ if $n = 0$ and is zero if $n \geq 1$.
 \subsection{Fractional Poisson process}

The fractional Poisson distribution, introduced by Repin and Saichev (2000) (see \cite{repin}), analyzes a PP with a stochastic intensity. Jumarie (2001) also contributed to similar concepts (see \cite{jumarie}). Laskin (2003) (see \cite{lask}) proposed a fractional generalization, known as the fractional Poisson process (fPP) denoted by $\{N_\beta(t)\}_{t\geq0}$. Its \textit{pmf} is given by
  $p_{\beta}(n,t)=\mathbb{P}[N_\beta(t)=n],$ it solves the following equation
 $$ \mathcal{D}_t^{\beta}p_{\beta}(n,t) = -\lambda\left[p_{\beta}(n,t)-p_{\beta}(n-1,t)\right], 0<\beta\leq 1,$$
 with $p_{\beta}(n,0)= 0$ if $n = 0$ and is zero if $n \geq 1$. Here $ \mathcal{D}_t^{\beta}$
 denotes the Dzhrbashyan-Caputo fractional derivative which is defined in \cite[Theorem 2.1]{KilSriTru06}).

The \textit{pmf} $p_{\beta}(n,t)=\mathbb{P}[N_\beta(t)=n]$ of the fPP is given by (see \cite{lask})
\begin{equation}
    \label{fpp-pmf}
 p_{\beta}(n,t) = \frac{(\lambda t^{\beta})^n}{n!}\sum_{k=0}^{\infty}\frac{(n+k)!}{k!}\frac{(-\lambda t^{\beta})^k}{\Gamma(\beta(k+n)+1)}.
\end{equation}
\subsection{Distributional properties of fPP} It is known that the mean and the variance of the fPP are  (see \cite{lask})
\begin{align}
\mathbb{E} [N_{\beta}(t)] &= qt^{\beta},\label{meanfpp}\\
\mbox{Var}[N_{\beta}(t)] &=qt^{\beta}\left[1+qt^{\beta}\left(\frac{\beta B(\beta, 1/2)}{2^{2\beta-1}}-1\right)\right],\label{varfpp}
\end{align}
where $q=\lambda/\Gamma(1+\beta)$ and $B(a,b)$ denotes the beta function. An alternative form for Var[$N_{\beta}(t)$] is given in \cite[eq. (2.8)]{beghinejp2009} as
\begin{equation*}
\mbox{Var}[N_{\beta}(t)]=q t^{\beta}+\frac{(\lambda t^{\beta})^{2}}{\beta}\left(\frac{1}{\Gamma(2\beta)}-\frac{1}{\beta\Gamma^{2}(\beta)}\right).
\end{equation*}

It was proved by  (\cite{fnbpfp}) that the one-dimensional distributions of the fPP are not infinite divisible. 

\subsection{Long memory and fPP}\label{sec:memory}
The significance and relevance of the long memory, also known as the long-range dependence is known in the literature. Its applications are well established  in several domains (see \cite{karag2004,Ding1993,Pagan1996,DouTaqqu2003,climate2006}). The main idea behind fPP comes from long-memory experimental data (see \cite{Cahoy2010}). The objective is to enhance the flexibility of the basic Poisson model by allowing inter-arrival times to follow non-exponential, heavy-tailed distributions and exhibit distinct scaling characteristics. It underlines the dependence of arrival of events in a counting process. Leonenko {\it et al.} (2014) (see \cite{LRD2014}) proved existence of  long memory property of the fPP. Authors in 
 \cite{biardapp,lrd2016} also showed that the increments of the fPP
have the LRD property.

\section{Parameter estimation using method of moments} \label{sec:mom}
In this section, we will derive the method of moments estimators (MoMEs) for the two parameters underlying the fPP namely, $\lambda$ and $\beta$. Unlike Cahoy (2011) (see \cite{Cahoy2010}), we will make use of the actual density function of the fPP, which takes the same form as in \eqref{fpp-pmf}. We opt for the following approach in order to find the MoMEs through simulations. 
\begin{itemize}
    \item We first simulate $n=1000$ independent sample paths of the fPP, for fixed values of the parameters.
    \item We then fix a threshold $t=300$, and find the number of occurrences of the fPP within each sample path, which are less than $t$. Let this be denoted by $n_t$.
\end{itemize}
The MoMEs can be derived by equating the first two theoretical moments, in \eqref{meanfpp} and \eqref{varfpp}, of the fPP, with the corresponding sample equivalents, which can be assumed to be $n_t$ and $n_t^2$ respectively. The two equations thus take the form,

\begin{align}
     E[N_{\beta}(t,\lambda)]&=qt^{\beta} = n_t  \\ \nonumber \\
     E[(N_{\beta}(t,\lambda))^2]&= qt^{\beta}\left[1+qt^{\beta}
     \left(\frac{\beta B(\beta,1/2)}{2^{2\beta-1}}-1\right)\right]
     +(qt^{\beta})^2 = n_t^2.
\end{align}
However, the above two equations cannot be solved analytically due to their complexities, and we hence resort to a numerical approach, namely the \textit{optim} function in $R$, to come up with the solutions. 

\begin{itemize}
    \item We find the MoMEs for each of the $1000$ sample paths generated earlier. Let these be denoted by $\lambda^*_j, \beta^*_j, j=1,2,...,1000$ respectively. 
    \item We then find the averages of the above $1000$ estimates, to find the final MoMEs, denoted by $\bar{\lambda^*},\bar{\beta^*}$ respectively. We also find the corresponding Bias and MSE for the MoMEs, by using the following formulas,
    \begin{align}
        \mbox{Bias}(\lambda^*,n)&=\frac{1}{1000}\sum_{i=1}^{1000}(\lambda^*-\lambda)\\ 
        \mbox{MSE}(\lambda^*,n)&=\frac{1}{1000}\sum_{i=1}^{1000}(\lambda^*-\lambda)^2,
    \end{align}
    and similarly for $\beta^*$.
\end{itemize}
  
Table \ref{table:mom} shows the mean MoMEs along with their Bias and MSEs. We fix the parameters as $\lambda=2, \beta=0.8$. One can see that under both the situations, the parameter estimates are close to the true values. The bias and MSEs are also close to 0, suggesting that the estimates seem unbiased. An interesting point to note is that the estimate for $\lambda$ tends to overestimate the true value, whereas that for $\beta$ tends to underestimate the actual value.

\begin{table}[H]
\centering
\begin{tabular}{c|ccc} 
 \hline
 && $\lambda=2, \beta=0.8$ &\\ \hline
 Estimate & MoME & Bias & MSE \\ \hline
 &&& \\
 $\bar{\lambda}^*$ & 2.0708 & 0.0708 &0.0187 \\ 
 &&&\\
 $\bar{\beta}^*$ & 0.7334 & -0.0665 & 0.0255 \\ \hline
\end{tabular}
\caption{Mean MoMEs along with their Bias and MSEs}
\label{table:mom}
\end{table}

\begin{figure}[H]
\includegraphics[width=7.5cm,height=5cm]{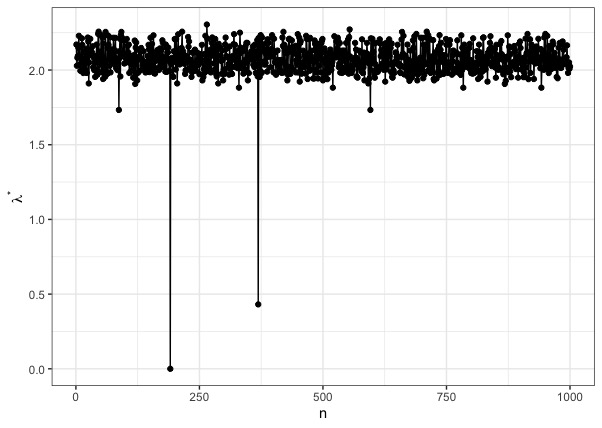}\hfill
\includegraphics[width=7.5cm,height=5cm]{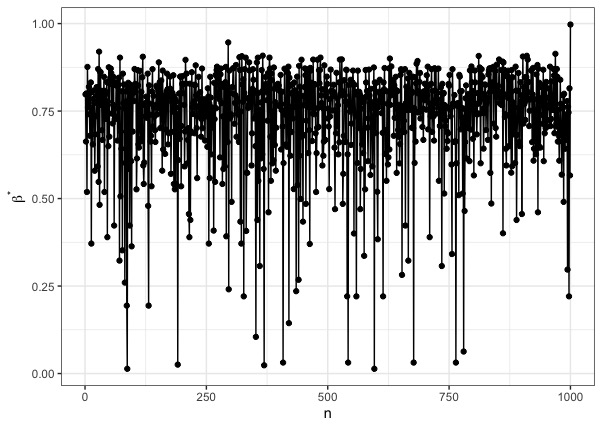}
\caption{Estimates for $\lambda$ and $\beta$ when the true values are $\lambda=2, \beta=0.8$.}
\label{figure:1}
\end{figure}

Figure \ref{figure:1} contains the trajectories of both the parameter estimates. One can see that in both the cases, the estimates are close to the actual values, apart from very few outliers.

\section{Parameter estimation - Maximum likelihood}\label{sec:MLE}
Assume for a fixed $m$, $n_{t1},n_{t2},...,n_{tm}$ be the observed data, where $n_{ti}, i=1,2,...,m$ denotes the number of occurrences of the fPP which are less than some threshold (say $t$), from the $i^{th}$ sample path. Then, the likelihood function can be written down as follows,
\begin{equation}
L(\lambda,\beta)=\prod_{i=1}^m \left\{\frac{(\lambda t^{\beta})^{n_{ti}}}{n_{ti}!}\sum_{k=0}^{\infty}\frac{(n_{ti}+k)!}{k!}\frac{(-\lambda t^{\beta})^k}{\Gamma(\beta (k+n_{ti})+1)}\right \},
\end{equation}
where $n_{ti}$ is the observed data as seen earlier. The MLEs have to be found out by applying the routine technique, that is to maximize the log-likelihood function with respect to the parameters. However as in the case of method of moments, the MLEs can't be solved analytically and hence a numerical approximation is required. We thus find the solutions using the \textit{nlm} package in $R$. We adopt the following approach to find the MLEs:
\begin{itemize}
    \item We first simulate $n=1000$ independent sample paths of the fPP, for fixed values of the parameter.
    \item As a particular instance, we randomly pick 50 sample paths, and find the number of occurrences which are less than the threshold ($t=300$). This denotes our first set of sample observations $n_{t1},n_{t2},...,n_{t50}$.
    \item We find the MLEs of the parameters using the above data, using the \textit{nlm} routine in $R$. We fix the value of $k$ as $49$, as it was observed that for $k>49$, the summand in the likelihood function is negligible and hence can be ignored.
    \item We repeat the above step 1000 times, to generate independent instances of the MLEs. We average the results to obtain the final MLEs. We also find the Bias and MSEs by using the formulas as seen earlier.
\end{itemize}

Table \ref{table:mle} contains the average MLEs along with their Bias and MSEs. We fix the parameter values as $\lambda=2, \beta=0.8$. The parameter estimates are seen to be close to the actual values, while the Bias and MSEs are close to 0, suggesting that the estimates are asymptotically unbiased.

\begin{table}[H]
\centering
\begin{tabular}{c|ccc} 
 \hline
 && $\lambda=2, \beta=0.8$ & \\ \hline
 Estimate & MLE & Bias & MSE  \\ \hline
 &&& \\
 $\bar{\hat{\lambda}}$ & 1.9699 & -0.0301 &0.0265  \\ 
 &&&\\
 $\bar{\hat{\beta}}$ & 0.7924 & -0.0075 & 0.0049 \\ \hline
\end{tabular}
\caption{Mean MLEs along with their Bias and MSEs}
\label{table:mle}
\end{table}

\begin{figure}[H]
\includegraphics[width=8cm,height=6cm]{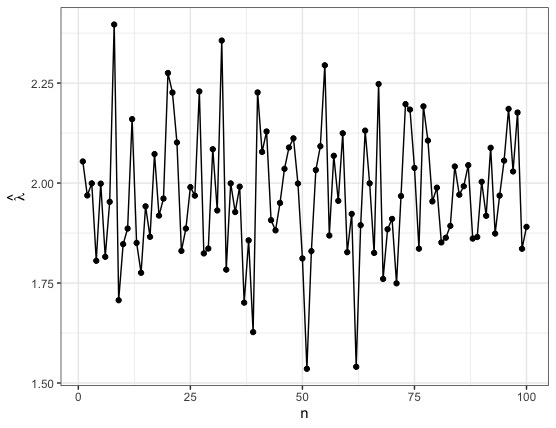}~
\includegraphics[width=8cm,height=6cm]{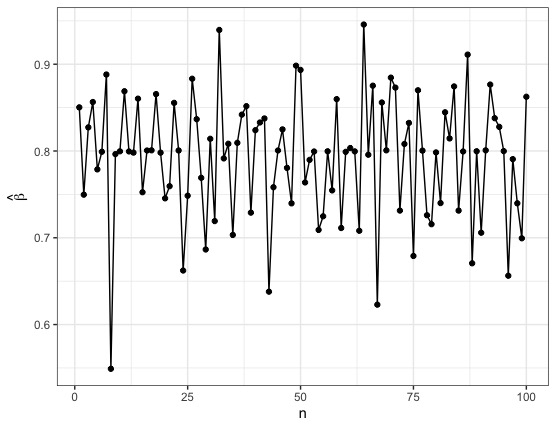}
\caption{Estimates for $\lambda$ and $\beta$ when the true values are $\lambda=2, \beta=0.8$}
\label{figure:2}
\end{figure}

Figure \ref{figure:2} gives the trajectories of both the parameter estimates. One can see that the estimates are located close to their actual values, with very few outliers.

Further, according to the regularity conditions stated in Rohatgi and Saleh (2000) (see \cite{rs2000}), one can define the distribution of the MLEs as a bivariate normal. The means become $(\lambda, \beta)$ and the variance covariance matrix is given by the inverse of the information matrix, which is denoted by $(I(\lambda,\beta)^{-1})$. The observed information matrix can be written as follows:
\begin{equation}
I((\lambda,\beta)|\lambda=\hat{\lambda},\beta=\hat{\beta})=\left.\begin{bmatrix}
\frac{-\partial^2\ln L(\lambda,\beta)}{\partial \lambda^2}&\frac{-\partial^2\ln L(\lambda,\beta)}{\partial \lambda\partial \beta}\\ \\

\frac{-\partial^2\ln L(\lambda,\beta)}{\partial \lambda\partial \beta}&\frac{-\partial^2\ln L(\lambda,\beta)}{\partial \beta^2}
\end{bmatrix}\right|_{\lambda=\hat{\lambda}, \beta=\hat{\beta}}.
\end{equation}

\begin{figure}[H]
\includegraphics[width=8cm,height=6cm]{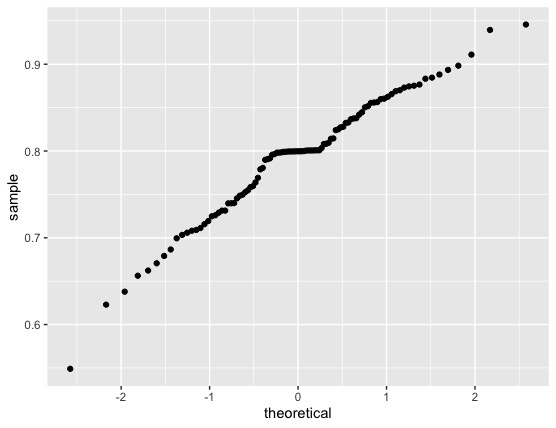}\hfill
\includegraphics[width=8cm,height=6cm]{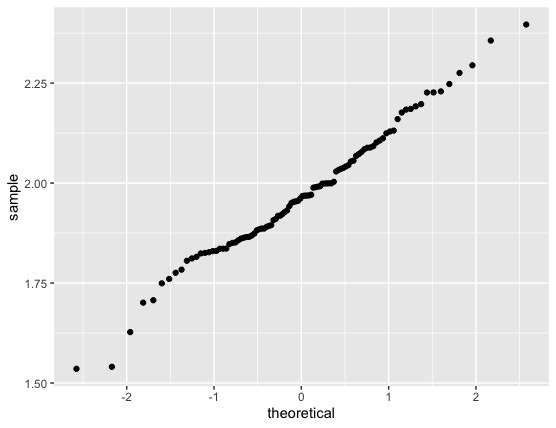}
\caption{Normal Q-Q plots for $\hat{\lambda}$ (left) and $\hat{\beta}$ (right).}
\label{figure:3}
\end{figure}

We also checked for asymptotic normality of our MLEs using the Kolmogorov-Smirnov test routine in $R$. The $p$-values were 0.4676 and 0.5806, suggesting that the MLEs indeed follow an asymptotic normal distribution. Same can be seen from Figure \ref{figure:3}, which gives the normal Q-Q plots for both the sets of MLEs.

\section{Practical data analysis}\label{sec:data}
Modeling wildfire occurrences across the globe has always been a challenge, mainly due to the complexity of such a data. The unpredictability of climate driven wildfires, along with their increasing size and coverage, makes it difficult to come up with a suitable model which can capture the behaviour of its patterns. In this paper, we will apply the fPP, suitably to model the wildfire occurrences across California, USA. Such a technique involving a simple parametric model for wildfire occurrences has not been concretely covered in literature to the best of our knowledge. A couple of attempts involving estimation using fPPs are by \cite{linhares2023fractional}, where the author has tried to model DNA sequences using a suitable fPP, or \cite{blender2015non}, where fPP has been used to model the occurrences of extreme mid-latitude cyclones during the boreal winter and summer seasons. \\\\
The wildfires in California have been of interest to many researchers with regards to modeling their occurrences or analyzing its other aspects. A few references coming from a variety of problem scenarios which one may follow are \cite{westerling2011climate,westerling2008climate}, where climate change and growth scenarios of wildfires are discussed, \cite{swain2021shorter}, where the authors postulate that a sharper rainy season amplifies the wildfire risks, \cite{aguilera2021wildfire}, where the impact of wildfire smoke on the respiratory health is discussed, or \cite{malik2021data}, where a data driven approach to model the wildfire risks is discussed.
\subsection{Dataset}
\begin{figure}[!ht]
\centering
\subfigure[2020]{
\includegraphics[width=.4\textwidth]{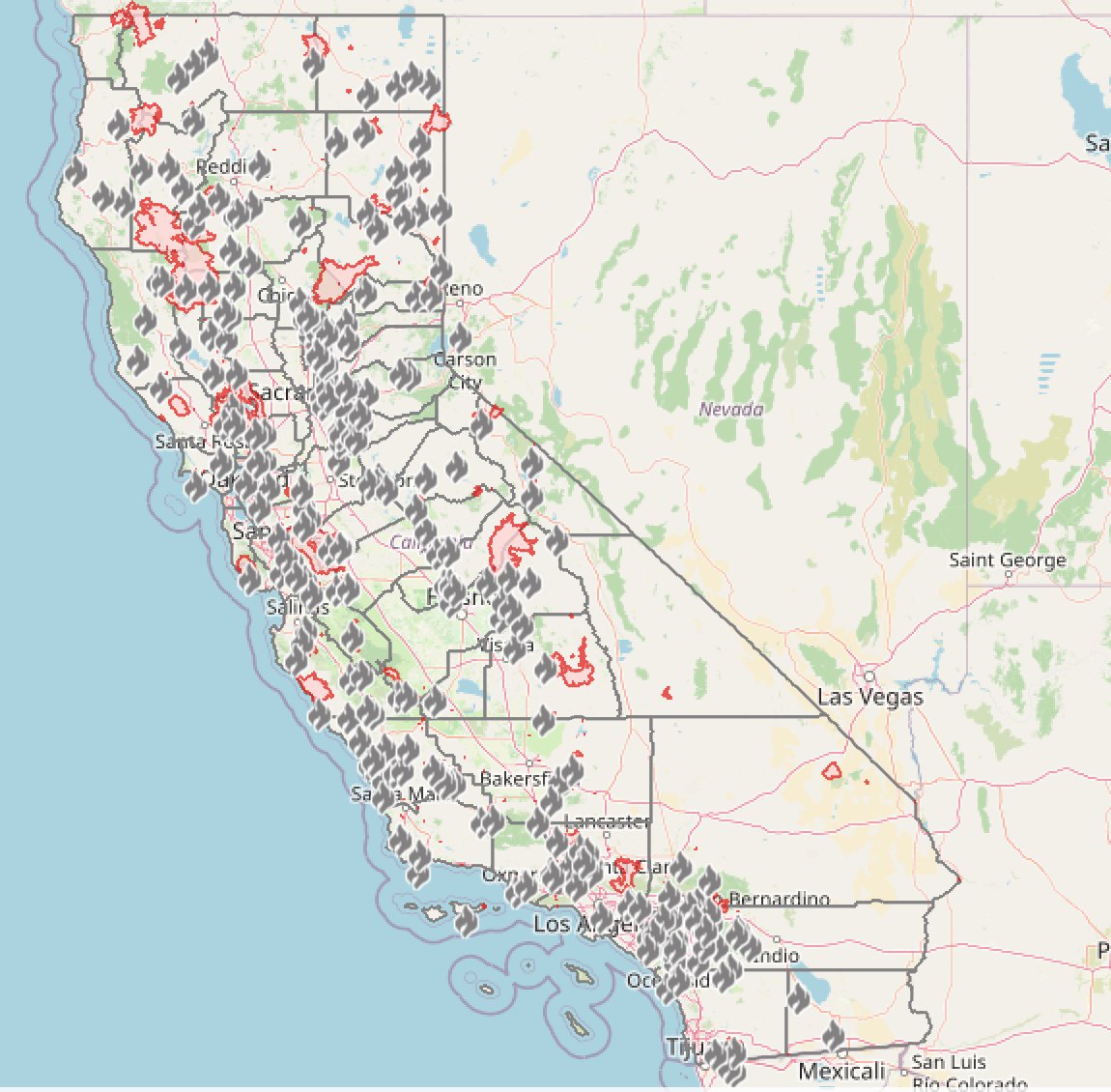}
}
\subfigure[2021]{
\includegraphics[width=.4\textwidth]{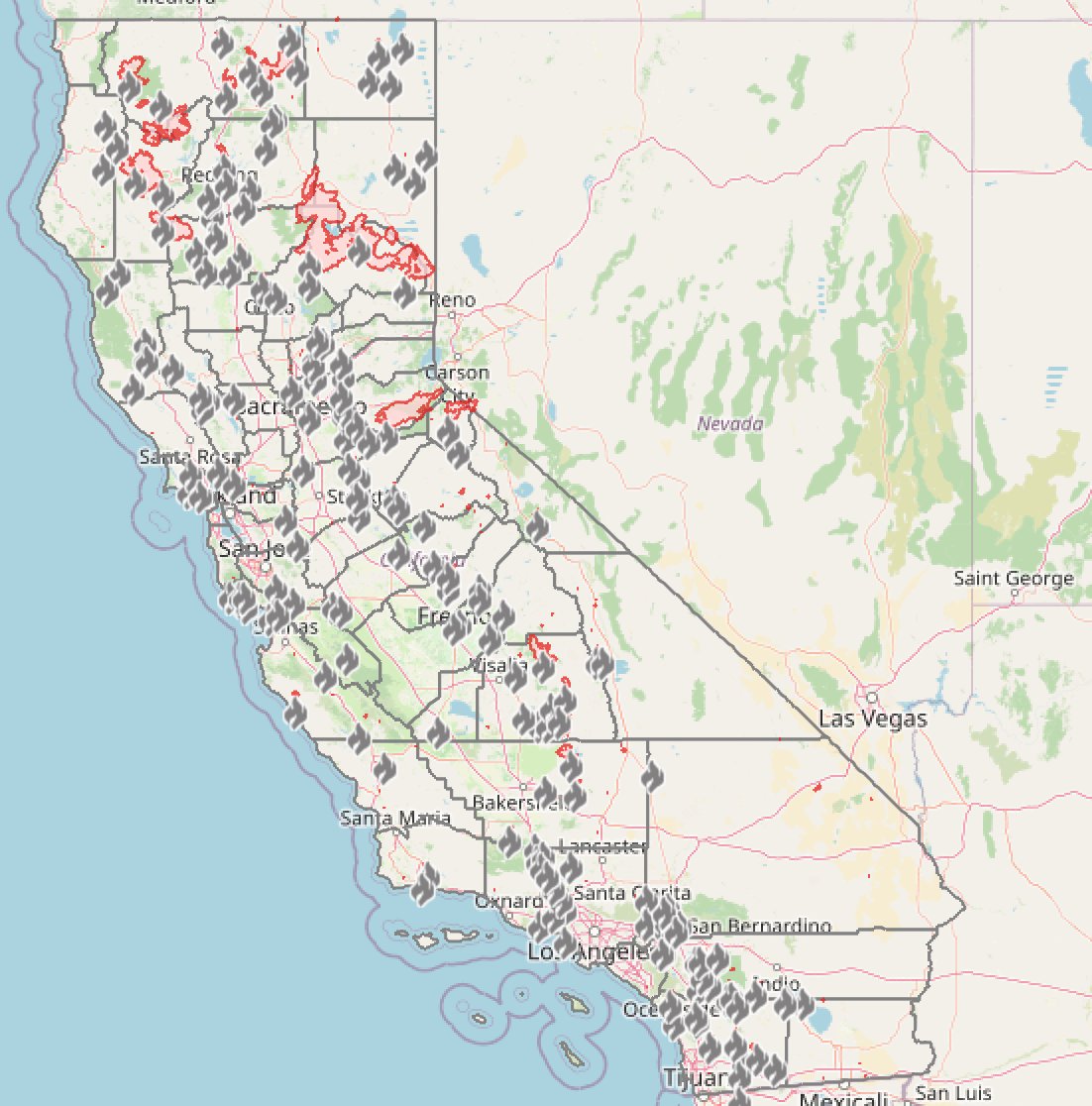}
}
\subfigure[2022]{
\includegraphics[width=.4\textwidth]{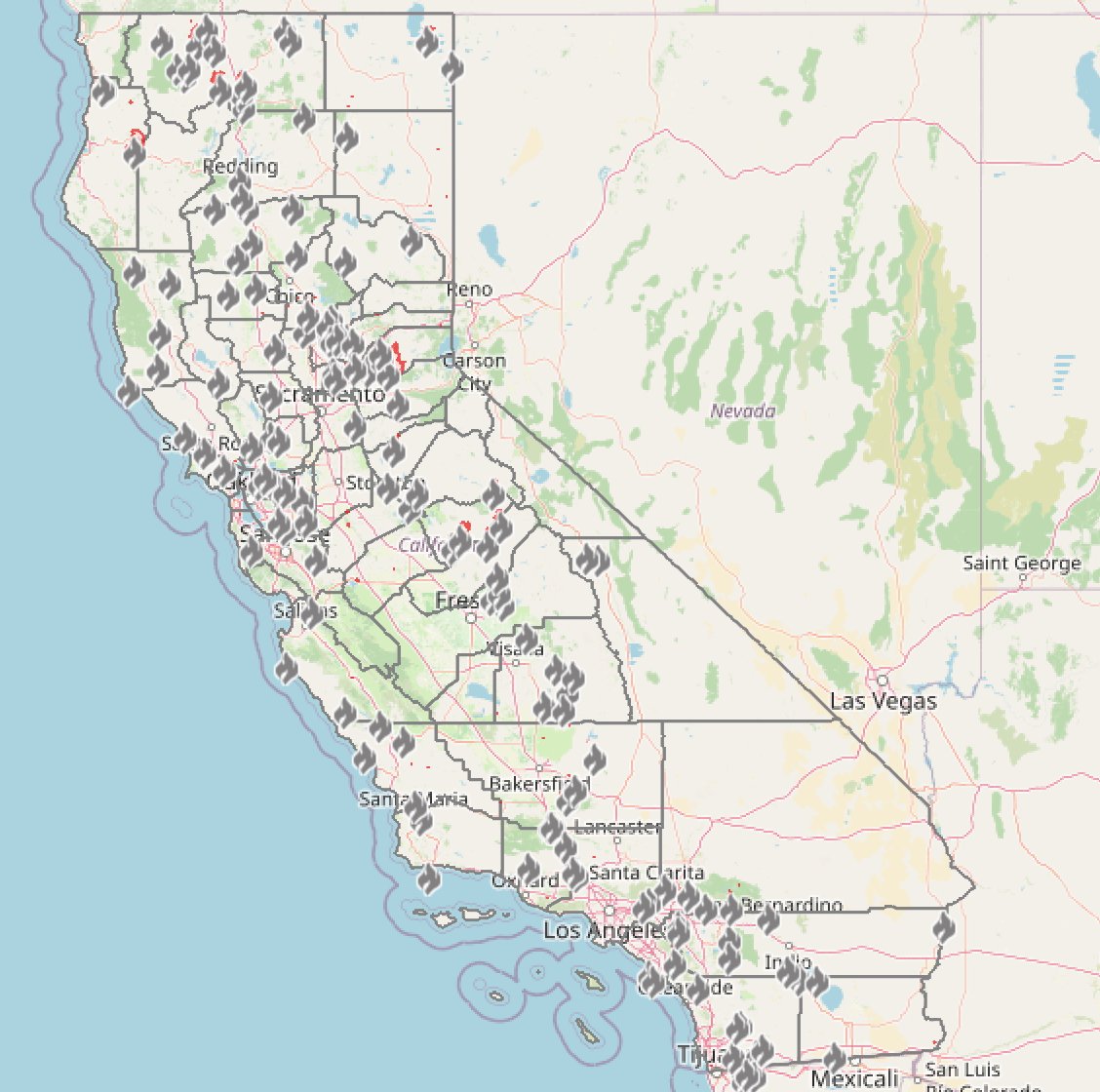}
}
\caption{California wildfire occurrences over three years. Provided by ``OpenStreetMap" under a CC-BY-SA license.}
\label{fig:open}
\end{figure}

As seen earlier, the dataset
analyzed in this paper is from the ``Storms Events Database", managed by the National Centers for Environmental Information, under the ``National Oceanic and Atmospheric Administration" (NOAA). The database contains records for different storms and other significant weather phenomena of sufficient intensity, occurring in the Unites States. It contains information on the event intensity, time of occurrence, number of injuries/deaths, damages to property etc. We will particularly look at the wildfire occurrences in California between June 2019 - April 2023\footnote{\href{https://www.ncdc.noaa.gov/stormevents/}{https://www.ncdc.noaa.gov/stormevents/}}. A few specifications are as follows: there were around 66 counties/zones which were affected. The number of days with wildfire occurrences were 146, out of which, 71 days involved either a death, injury or damage to property or crops. The website also provides detailed information of each event, describing the coverage, damaged area, an episode narrative, an event narrative etc. For a perspective, Figure \ref{fig:open} contains plots of wildfire events across California, between 2020-2022. The year 2020 was more severe, witnessing larger and devastating fires being reported. The maps are taken from the Cal-Fire's website (\href{https://www.fire.ca.gov/incidents/}{https://www.fire.ca.gov/incidents/}), and are provided by ``OpenStreetMap" contributors under a CC-BY-SA license. 

\begin{figure}[!ht]
\subfigure{
\includegraphics[width=8cm,height=5cm]{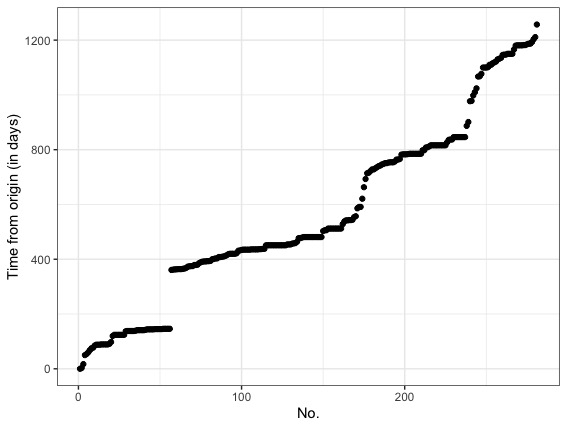}}\hfill
\subfigure{\includegraphics[width=8cm,height=5cm]{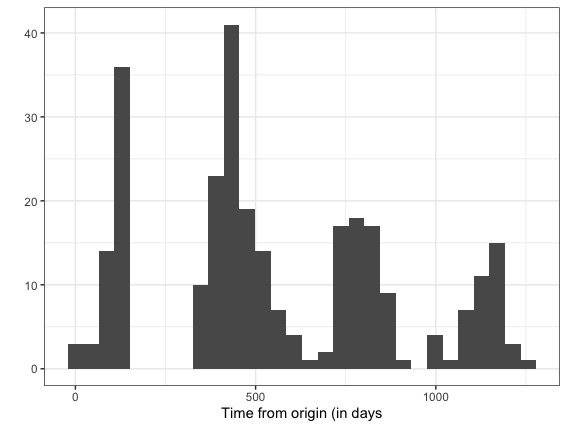}}
\caption{Scatterplot (left) and histogram (right) of the times from origin, in days.}
\label{figure:4}
\end{figure}

Now treating the first wildfire occurrence as the origin, we find the time until the next wildfire occurrence (in days) from the origin, and record that as an observation. For example, the first three occurrences were on June 8th, June 11th and June 25th, 2019. Hence, the first three data points are 0, 3 and 17 respectively. Figure \ref{figure:4} contains a preliminary scatterplot and a histogram of the times from origin. One can see the clear jumps and irregular occurrences throughout the time range. One can thus sense that a suitable model will be far from a regular PP, where the occurrences and jumps are consistent throughout.

\subsection{Estimation}
We now try to fit a suitable fPP on the wildfire occurrences data, by estimating the parameters using the method of moments approach. We take the following steps in this regard: 
\begin{itemize}
    \item We fix $t=200$ days, and find the number of occurrences less than $t$.
    \item Thus we get a single sample observation $N_\beta(t)$.
    \item Using the method of moments equations, (5) and (6), we estimate the underlying parameters $\lambda, \beta$.
\end{itemize}

\begin{figure}[!ht]
\centering
\includegraphics[width=14cm,height=12cm]{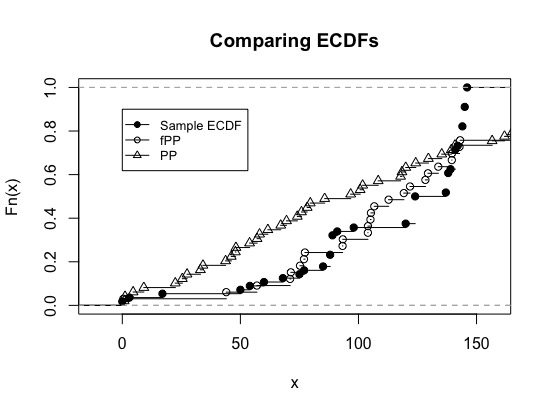}
\caption{Sample ECDFs of the actual data and the two processes}
\label{figure:ecdf}
\end{figure}

Using the above steps, it was observed that $N_\beta(t)=56$, while the estimators are $\lambda^*=0.69, \beta^*=0.8$. Now as a comparison, we also try to fit a regular PP $N(t)$ on the same data, and the estimated parameter comes out to be $\lambda^*=0.28$. Figure \ref{figure:ecdf} contains the sample ECDF plots of the actual data and the two processes which we have fit. Once can clearly see that the fPP is far better than the PP, in capturing the irregular jumps among the wildfire occurrences. We also run a Kolmogorov-Smirnov (K-S) test to check for the suitability of the fit. The $p$-values were observed to be 0.1761 and 0.00034 respectively for the fPP and PP fits. This supports the claim that the fPP fits better than the PP, and hence can be applied to model wildfire occurrences suitably.

\subsection{Discussion} In the previous subsection, the estimation has been carried out using the method of moments approach instead of the MLE approach. This is because a real data generates a single set of observations, from which one can only find a single value of the number of occurrences $N_\beta(t)$ which are less than a fixed value $t$. Finding the MLE is hence not possible from a single sample value. A small workaround to be able to still use the MLE approach is to generate copies of pseudo-real sets of observations from the observed value of $N_\beta(t)$. The method of moments approach however proves to be easy and robust in this case. It can also be noted that estimated parameter $\beta^*=0.8$ proves existence of long-memory in the  wildfire data (see Section \ref{sec:memory}).

\section{Prediction of wildfires with the fPP model}\label{sec:pred}
As seen in the last section, the fPP model turned out to be better than a standard PP in capturing the wildfires across California during June 2019 - April 2023. In this section, using the same practical dataset, we try to predict the occurrences of wildfires using the proposed fPP model. We will then compare these predictions with those obtained from a PP. In order to do so, we will make use of the inter arrival times of both the processes to generate the next occurrence of a wildfire. The inter arrival time for a PP is simply an $Exponential$ distribution with parameter $\lambda$, whereas that for a fPP is given by (see \cite{Kanter1975})
\[T=\frac{|\log(U_1)|^{1/\beta}}{\lambda^{1/\beta}}\frac{\sin(\beta \pi U_2)[\sin((1-\beta)\pi U_2)]^{1/\beta-1}}{[\sin(\pi U_2)]^{1/\beta}|\log(U_3)|^{1/\beta-1}},\]
where $U_1,U_2,U_3$ are independent Uniform random variables between $(0,1)$. We adopt the following stepwise approach to carry out the prediction and comparison:
\begin{itemize}
    \item We fix $t=200$, which leads to the number of wildfire occurrences being 56, as seen from the last section.
    \item The estimated parameters are $\lambda^*=0.69, \beta^*=0.8$ for the fPP and $\lambda^*=0.28$ for the PP.
    \item Using the above estimated parameters, we find the inter arrival times in days, for 10 future wildfire occurrences for both the processes.
    \item Using the inter arrival times, we then find the time points of actual wildfire occurrences for both the processes.
    \item We then compare these time points with those coming from the actual data using a few well known metrics such as MSE and the mean absolute deviation (MAD).
\end{itemize}
Now assuming the first wildfire occurrence on June 8th, 2019 as the origin, the following table shows the predicted inter arrival times (IA) after the $56^{th}$ wildfire occurrence, and the time of occurrence (TO) in days, from the origin, for both the processes. The table also contains times of actual occurrences in days, from the origin, in the last row.

\begin{table}[H]
\centering
\begin{tabular}{c|cccccccccc} 
  & $t_1$ & $t_2$ & $t_3$&$t_4$&$t_5$&$t_6$&$t_7$&$t_8$&$t_9$&$t_{10}$ \\ \hline
 &&& \\
 fPP-IA & 189.56 & 11.74 & 6.35& 8.04& 4.81& 5.29& 9.47&8.24& 14.84& 6.98  \\ 
 &&&\\
 PP-IA & 3.31 & 3.43 & 3.66 & 3.53 & 3.49 & 3.83 & 3.43 & 3.51 & 3.48 & 3.50 \\
 &&&\\
 fPP-TO & 335.56 & 347.30 & 353.65 & 361.69 & 366.5 & 371.79 & 381.26 & 389.5 & 404.34 & 411.29\\
 &&&\\
 PP-TO & 149.52 & 153.08 & 156.78 & 160.47 & 163.93 & 167.57 & 171.15 & 174.89 & 178.45 & 182.01 \\ 
 &&&\\
 Actual TO & 361 & 362 & 363 & 363 & 364 & 364 & 364 & 365 & 366 & 368\\ \hline
\end{tabular}
\caption{Predicted inter arrival times, predicted time of occurrences and the actual time of occurrences in days}
\label{table:1}
\end{table}
From the above table one can clearly note that the PP generates inter arrival times at a consistent rate, which does not reflect the true behaviour of the wildfires. On the other hand, the fPP is able to generate erratic inter arrival times, which are comparable to the actual time of occurrences. As a formal comparison, the below table gives the values of MSE (as defined in Eq. (9)) and MAD for both the processes of the predicted time of occurrences, where the MAD is as per the below formula.

\begin{align}
    MAD(\lambda^*,n)&=\frac{1}{n}\sum_{i=1}^{n}|\lambda^*-\lambda|
\end{align}

\begin{table}[H]
\centering
\begin{tabular}{c|ccc} 
  & $n$ & MSE & MAD  \\ \hline
 &&& \\
 fPP & 10 & 526.14 & 18.45  \\ 
 &&&\\
 PP & 10 & 39562.1 & 198.72 \\ \hline
\end{tabular}
\caption{MSE and MAD for the two processes}
\label{table:1}
\end{table}
As one can see from the above table, fPP is much better than the usual PP to model the wildfire occurrences, both in terms of MSE and MAD. One can observe more than $90\%$ reduction in the errors for fPP as compared to PP. The below figure gives a visual comparison between the predicted wildfire occurrences of the two processes, along with the actual data. One can again see that the fPP is able to predict the wildfire occurrences much closely as compared to the PP model.

\begin{figure}[H]
\centering
\includegraphics[width=12cm,height=8cm]{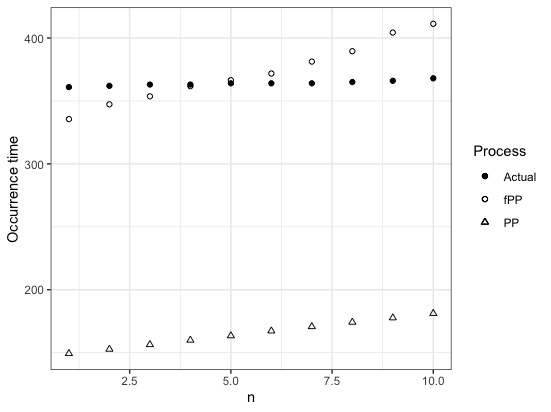}
\caption{Predicted time of occurrences with the actual data}
\label{figure:5}
\end{figure}

\subsection*{Conclusion and future work} In this paper, we have focused on statistical estimation of the parameters of a fPP using both the method of moments approach and the maximum likelihood approach. Estimation using the probability density function of the fPP is novel, which has not been covered in the literature yet. We further modeled wildfire occurrences across California, USA, using an appropriate estimated fPP. The fPP was seen to be much better than the usual PP in this regard. Prediction was also carried out using the estimated fPP model, and compared with the PP. There are the two main takeaways from this work:
\begin{itemize}
    \item The estimates of fractional parameter $\beta^*$ for the fPP for wildfire data turns out to be $0.8$, which shows that wildfire events have long-memory and the events of wildfire events are dependent. This is a clear departure from the PP  model and will certainly call for more nuanced investigation in this field.
    \item The reduction in prediction error is close to 90\%, which reinforces the above finding and strongly endorses the usefulness of the fPP model in this scenario. We believe that these prediction methods will greatly improve understanding of wildfire events among various stakeholders, such as, insurance companies, local administration, firefighting and forest departments.\end{itemize}
As a future outlook, one may look to predict the amount and extent of damage (property or crop) due to wildfires, using a suitable compound Poisson process structure. One can also look at the application of the fPP model in other disaster events and in other terrains. This will be highly beneficial for the local government agencies to plan out a strategy accordingly.

\section*{Declarations}
\subsection*{Conflict of interest}
On behalf of both the authors, the corresponding authors declares that there is no conflict of interest.

\ifx 
\textcolor{red}{we may highlight following points}
\begin{enumerate}
    \item Useful prediction from application perspective
    \item Memory of decay
    \item Prediction table and compare it wrt Poisson prediction
    \item Highlight difference and usefulness in decision making. Underline cost associated to it. 
\end{enumerate}

\bibliographystyle{abbrv}

\bibliography{researchbib}

\fi 
\def\cprime{$'$}

\end{spacing}

\end{document}